\documentclass[prb,twocolumn,preprintnumbers,superscriptaddress]{revtex4}

\usepackage{graphicx}
\usepackage{amssymb,amsmath}
\usepackage{dcolumn}
\usepackage{bm}
\usepackage{color}
\usepackage{enumerate}

\begin{document}

\title{Adiabatic preparation of a cold exciton condensate}

\author{V. Shahnazaryan}
\affiliation{Science Institute, University of Iceland, Dunhagi-3, IS-107, Reykjavik, Iceland}
\affiliation{Institute of Mathematics and High Technologies, Russian-Armenian (Slavonic) University, Hovsep Emin 123, 0051, Yerevan, Armenia}

\author{O. Kyriienko}
\affiliation{Science Institute, University of Iceland, Dunhagi-3, IS-107, Reykjavik, Iceland}
\affiliation{Division of Physics and Applied Physics, Nanyang Technological University 637371, Singapore}
\affiliation{QUANTOP, Danish Quantum Optics Center, Niels Bohr Institute, University of Copenhagen, Blegdamsvej 17, DK-2100 Copenhagen, Denmark}

\author{I. A. Shelykh}
\affiliation{Science Institute, University of Iceland, Dunhagi-3, IS-107, Reykjavik, Iceland}
\affiliation{Division of Physics and Applied Physics, Nanyang Technological University 637371, Singapore}

\date{\today}

\begin{abstract}
We propose a scheme for the controllable preparation of a cold indirect exciton condensate using dipolaritonic setup with an optical pumping. Dipolaritons are bosonic quasiparticles which arise from the coupling between cavity photon (C), direct exciton (DX), and indirect exciton (IX) modes, and appear in a double quantum well embedded in a semiconductor microcavity. Controlling the detuning between modes of the system, the limiting cases of exciton-polaritons and indirect excitons can be realized. Our protocol relies on the initial preparation of an exciton polariton condensate for the far blue-detuned IX mode, with its subsequent adiabatic transformation to an indirect exciton condensate by lowering IX energy via applied electric field. The following allows for generation of a spatially localized cold exciton gas, on the contrary to currently used methods, where IX cloud appears due to diffusion of carriers from spatially separated electron- and hole-rich areas.
\end{abstract}

\maketitle

\section{Introduction}

Experimental observation of Bose-Einstein condensation
(BEC)\cite{Pitaevskii2006} in a cold atomic gas system represents a
major breakthrough of contemporary condensed matter
physics.\cite{Anderson1995,Davis1995} However, while being a perfect
testbed for studies of macroscopically coherent phenomena, the
requirement of ultralow temperature ($< 1~\mu$K) of an atomic gas
restricts its possible applications. Fortunately, this requirement
can be removed in solid state setups, where condensation of excitons
in semiconductor structures was theoretically predicted
\cite{Keldysh1965,Blatt1962} for considerably higher temperatures
($\approx 1$~K). While unambiguous evidence of exciton condensation
in a bulk and monolayer semiconductor structure is still lacking due
to technological difficulties,\cite{SnokeBook} several modifications
of conventional excitonic system have shown promising results.

For instance, exciton polaritons, being a hybrid quasiparticles of
coupled excitons in semiconductor quantum well and confined cavity
photon,\cite{Kavokin2007,Deng2010} represent a useful setup for
studies of bosonic phenomena due to large decoherence times and
simplicity of experimental characterization using optical
techniques. The extremely small mass of these quasiparticles allowed
to rise the typical critical temperature of condensation to tens of
Kelvins range,\cite{Kasprzak2006,Balili2007,Wertz2010,Fischer2014}
and in certain cases up to room
temperature.\cite{Baumberg2008,Plumhof2014} With experimental
advances the macroscopically coherent polariton gas is now routinely
observed in numerous system, working both with incoherent
optical\cite{Kasprzak2006,Balili2007} and
electrical\cite{Schneider2013,Bhattacharya2013} pump. The unique
properties of polaritons enabled observation of
solitons,\cite{Sich2013,Sich2014,Amo2011} quantized
vortices,\cite{Lagoudakis2008,Manni2012} polarization
effects,\cite{Leyder2007,Tosi2011,Adrados2010,Shelykh2010} and are
considered as a platform for optical
computing.\cite{Espinosa2013,Liew2010} At the same time, their short
lifetime coming from cavity photon mode leakage does not allow for
proper thermalization of particles, and essentially restricts the
system to driven-dissipative non-equilibrium
behavior.\cite{CarusottoRev}

Another bosonic system where long-range coherence was observed is
represented by spatially indirect excitons---composite electron-hole
objects formed in a double quantum
well.\cite{High2012,Butov2004a,Butov2007} An intense research in
this area revealed various phenomena, including ring
fragmentation,\cite{Butov2002,Snoke2002,Dubin2012} polarization
patterns and spin currents,\cite{High2013,Leonard2009,Violante2014}
electrical\cite{Winbow2011,Andreakou2014} and
acoustic\cite{Lazic2014} routing etc. Given the reduced spatial
overlap of electron and hole wavefunctions,\cite{Sivalertporn2012}
which causes long lifetime of indirect exciton ($> 100$
ns),\cite{Butov2001} the efficient thermalization of particles can
be achieved.\cite{Butov2002} Simultaneously, the reduced
photon-exciton interaction constant complicates the optical
generation and characterization of indirect exciton
gas.\cite{Nalitov2014a} In particular, the common scheme for IX
cloud preparation includes optical generation of holes in one
quantum well and electrical injection of electron in the adjacent
well.\cite{Butov2007} This largely impedes the spatial control of
cold exciton gas, and for instance is believed to cause a
fragmentation of an indirect exciton
cloud.\cite{Butov2004b,Rapaport2004}

Recently, the possibility to unite the subjects of exciton-polaritons and indirect excitons was attained in the system of dipolaritons.\cite{Cristofolini2012,Christmann2011} Being hybrid quasiparticles consisting of cavity photon (C), direct exciton (DX), and indirect exciton (IX), they share desirable properties of both
polaritons and cold excitons, including enhanced interparticle interactions, enlarged lifetime, and improved optical control. The system of dipolaritons was proposed to serve as an efficient terahertz emitter,\cite{Kyriienko2013,Kristinsson2013,Kristinsson2014} tunable single-photon source,\cite{Kyriienko2014} and polarization switcher.\cite{Nalitov2014b}

In this paper we propose a scheme for optical generation of a cold
indirect exciton condensate using a dipolariton setup. At the first
step, it requires an initial preparation of an exciton polariton
condensate using the incoherent optical pump, with indirect exciton
mode lying high in energy for zero applied electric field $F$. By
lowering IX energy with an increase of field $F$, the lower
dipolariton state experiences adiabatic Landau-Zener
transition,\cite{Sinitsyn2013,Liew2012} converting to indirect
excitonic state with high fidelity. The following allows for
generation of a spatially localized cold exciton gas, on the
contrary to currently used methods, where IX cloud appears due to
diffusion of carriers from spatially separated electron- and
hole-rich areas.


\section{The model}
\begin{figure}[t]
\includegraphics[width=1.0\linewidth]{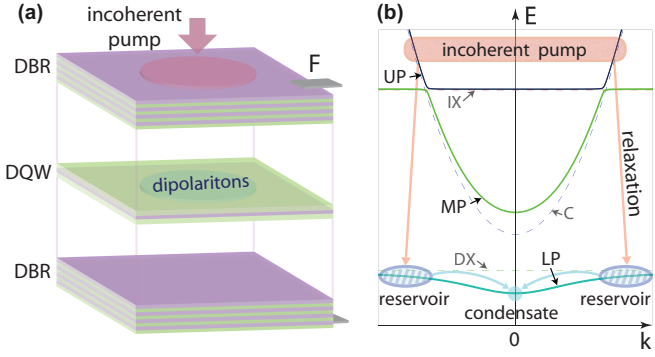}
\caption{(Color online) (a): Sketch of the dipolaritonic system, representing double quantum well (DQW) structure embedded in a microcavity formed by distributed Bragg reflectors (DBRs). (b): Schematic representation of dispersion of lower (LP), middle (MP), and upper (UP) dipolaritons. Incoherent optical pump creates carriers at high energy, which relax to LP reservoir, and consequently scatter to a macroscopically coherent ground state.}
\label{Fig1}
\end{figure}

The considered structure consists of two quantum wells separated by
a thin barrier, allowing electron to tunnel between the wells. This
double quantum well (DQW) system is placed in an optical
microcavity, providing the existence of a cavity photon mode
strongly coupled to a direct exciton mode, while an indirect exciton
mode remains decoupled from the light mode [Fig. \ref{Fig1}(a)]. An
indirect exciton instead is coupled to direct exciton due to
coherent tunnel coupling between the QWs. The bias is applied to the
heterostructure in the growth direction, which allows to tune the
energy of an indirect exciton and thus the tunneling efficiency.

The Hamiltonian of dipolariton system can be written in the general
form $\hat{H}=\hat{H}_{\rm{coh}}+\hat{H}_{\rm{dec}}$, where coherent
and decoherent processes are separated. The coherent part of
Hamiltonian reads:
\begin{equation}
    \begin{aligned}
\hat{H}_{\rm{coh}}&=\hbar\omega_C \hat{a}^\dag \hat{a}+\hbar\omega_{DX} \hat{b}^\dag \hat{b}+\hbar\omega_{IX}(t) \hat{c}^\dag \hat{c}\\
&+\frac{\hbar\Omega}{2} (\hat{a}^\dag \hat{b}+\hat{b}^\dag \hat{a})-\frac{\hbar J}2 (\hat{b}^\dag \hat{c}+\hat{c}^\dag \hat{b}),\\
    \end{aligned}
\label{H_coh}
\end{equation}
where $\hat{a}^\dag$, $\hat{b}^\dag$ and $\hat{c}^\dag$ are creation operators of cavity photons, direct excitons, and indirect excitons, respectively. First three terms in Eq. (\ref{H_coh}) correspond to energies of photon ($\hbar \omega_C$), direct exciton ($\hbar \omega_{DX}$) and indirect exciton ($\hbar \omega_{IX}$) modes. The next two terms describe the direct exciton-cavity photon Rabi splitting $\hbar\Omega$ and the direct-indirect exciton tunneling splitting $\hbar J$. Here, we emphasize that an indirect exciton energy is time dependent, and depends linearly on the time-varied applied electric field $F(t)$, $\hbar \omega_{IX}(t) = \hbar \omega_{IX}^{(0)} - e L F(t)$. This expression hold for narrow QW heterostructure; $\hbar\omega_{IX}^{(0)}$ is an energy of indirect exciton at zero bias; $L$ is a distance between centers of QWs; $e$ denotes electron charge.

For the strong coupling case where intermode couplings $\Omega$ and
$J$ overcome decay (or broadening) of the modes, the eigenstates of
Hamiltonian (\ref{H_coh}) correspond to lower (LP), middle (MP), and
upper (UP) dipolariton states, being coherent superpositions of
original C, DX, and IX states. The sketch of dipolariton dispersions
is shown in Fig. \ref{Fig1}(b) for largely blue-detuned indirect
exciton mode, and positive detuning of cavity photon with respect to
direct exciton, $\omega_{C}-\omega_{DX} > 0$.

The initial feeding source of the dipolariton system is represented
by laser tuned to high energies [Fig. \ref{Fig1}(b)]. It excites
free electrons and holes, which relax to low energies, forming
excitons with large wave vectors, commonly referred as reservoir
states.\cite{Wouters2007} During this fast relaxation process the
initial phase of the laser is fully lost, and incoherent reservoir
of the direct excitons is created. Next, the excitons from reservoir
can scatter due to exciton-phonon interactions towards ground state
at zero wave vector of lower dipolariton mode, where they form a
macroscopically occupied coherent state. The following excitation
scheme is commonly used in conventional polaritonic
setups,\cite{Kasprzak2006} where nonequilibrium condensation of
polaritons under incoherent pumping conditions was observed.

To describe incoherent processes related to phonon-assisted
scattering of particles from reservoir to the ground state, we
introduce the exciton-phonon interaction Hamiltonian
$\hat{H}_{\rm{dec}}=\hat{H}^+ +\hat{H}^-$, where
\begin{equation}
\hat{H}^+=D_{ph}\sum_{k}\hat{b}^\dag\hat{r}_k\hat{d}^\dag_k; \qquad \hat{H}^-=D_{ph}\sum_{k}\hat{b}\hat{r}^\dag_k\hat{d}_k,
\label{H_+}
\end{equation}
correspond to processes with emission ($\hat{d}_{k}^\dag$) and
absorption ($\hat{d}_{k}$) of phonons with wave vector $k$.
$\hat{r}_{k}^\dag$ and $\hat{r}_{k}$ are creation and annihilation
operators for reservoir states. $D_{ph}$ denotes exciton-phonon
interaction constant.

In order to take into account the decoherence caused by a finite
lifetime of the modes and interaction with reservoir, one can use
the Lindblad master equation for the density matrix $\rho$,
\begin{equation}
\frac{\partial \rho}{\partial t}=\frac{i}{\hbar}[\rho,\hat{H}]+\hat{\mathcal{L}}^{\rm{(dis)}}\rho +\hat{\mathcal{L}}^{\rm{(th)}}\rho,
\end{equation}
where $\hat{\mathcal{L}}^{\rm{(dis)}}$ is Lindblad superoperator
having the form $\hat{\mathcal{L}}^{\rm{(dis)}}\rho =\sum_i \gamma_i
(\hat{a}_i\rho \hat{a}_i^\dag-\{\hat{a}_i^\dag \hat{a}_i,\rho \}/2)$
with $\hat{a}_i= \hat{a}, \hat{b}, \hat{c}$ and $\gamma_j=1/\tau_j$
($j=C,DX,IX$) being damping rates of the modes.\cite{Kyriienko2013}
The term $\hat{\mathcal{L}}^{\rm{(th)}}\rho$ corresponds to
phonon-assisted processes accounted using Born-Markov
approximation.\cite{SM}
\begin{figure}[t]
\includegraphics[width=1.0\linewidth]{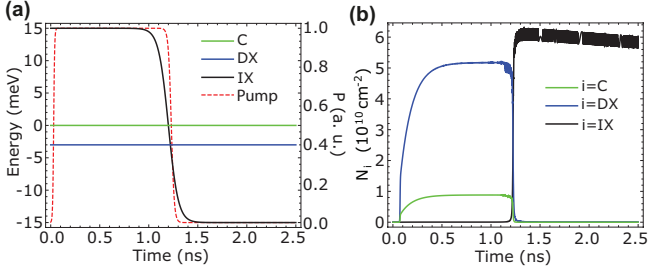}
\caption{(Color online) (a): Time dependence of energies of the modes. The bias applied to the system causes linear decrease of IX energy in $t=1$ ns to $t=1.5$ ns window, up to far red-detuned value. The dashed red line corresponds to time dependence of the pump intensity (in arbitrary units). (b): The evolution of occupations of the modes, being $N_C = |\langle \hat{a}\rangle|^2$, $N_{DX} = |\langle \hat{b} \rangle|^2$, and $N_{IX} = |\langle \hat{c} \rangle|^2$. At the first stage ($t<1$ ns) the formation of polariton condensate takes place. Next, continuous change of an applied bias drives the system through an avoided crossing, leading to the transfer of polariton occupation to an indirect exciton mode.}
\label{Fig2}
\end{figure}

Since we are interested in a large number of particles, we can apply
the mean field approximation when time dynamics of the system can be
defined by equations for mean fields given by $\partial
\langle\hat{a}_i\rangle / \partial t = \mathrm{Tr}(\hat{a}_i
\partial \rho/\partial t)$, where $\hat{a}_i= \hat{a}, \hat{b},
\hat{c}$.
%
%

Using the straightforward algebra we get equations of motions for
the cavity photon, direct exciton, and indirect fields coupled to
reservoir:\cite{SM}
\begin{equation}
\frac{\partial \langle\hat{a}\rangle}{\partial t}=-i\omega_C \langle\hat{a}\rangle-i\frac{\Omega}{2} \langle\hat{b}\rangle-\frac{\gamma_C}{2}\langle\hat{a}\rangle,
\label{eq_a}
\end{equation}
\begin{align}
\label{eq_b}
\frac{\partial \langle\hat{b}\rangle}{\partial t}=&-i\omega_{DX} \langle\hat{b}\rangle-\frac{\Omega}{2} \langle\hat{a}\rangle+i\frac{J}{2}\langle\hat{c}\rangle -\frac{\gamma_{DX}}{2}\langle\hat{b}\rangle \notag \\
&+\frac{W}{2}\langle\hat{b}\rangle \left(N_R-N_{ph}\right),
\end{align}
\begin{equation}
\label{eq_c}
\frac{\partial \langle\hat{c}\rangle}{\partial t}=-i\omega_{IX}(t) \langle\hat{c}\rangle+i\frac{J}{2}\langle\hat{b}\rangle -\frac{\gamma_{IX}}{2}\langle\hat{c}\rangle,
\end{equation}
\begin{align}
\label{eq_n_R} \frac{\partial N_R}{\partial t}= P(t)
-\gamma_{R} N_R -W|\langle\hat{b}\rangle|^2
\left(N_R-N_{ph}\right),
\end{align}
where $N_R=\sum_{k}n^R_k\equiv \sum_{k}\langle\hat{r}_k^\dag
\hat{r}_k\rangle$ denotes the full occupancy of the reservoir and
$N_{ph}=\sum_{k}n_k^{ph}$ corresponds to the total number of the
phonons defined by the temperature of the sample.
$W=2\delta_{R}D_{ph}^2$ corresponds to the scattering rate of
reservoir particles to macroscopically coherent state, where
$\delta_{R}$ is the inverse broadening of exciton states divided by
$\hbar^2$. The term $P(t)$ in Eq. (\ref{eq_n_R}) corresponds to the incoherent pump of reservoir states, which is typically given by the Lindblad type operator written in the form:
\begin{equation}
\hat{\mathcal{L}}^{\rm{(pump)}} \rho= \sum_k P_k(t)\Big(\hat{r}_k\rho\hat{r}_k^\dagger +\hat{r}^\dagger_k\rho\hat{r}_k- \hat{r}_k^\dagger\hat{r}_k\rho -\rho\hat{r}_k\hat{r}_k^\dagger\Big),
\end{equation}
where $|P_k(t)|^2$ denotes the intensity of incoherent pump of single reservoir state with in-plane wave vector $k$, and total pump is defined as $P(t)=\sum_{k} P_{k}(t)$. Writing the system of the kinetic equations we assumed large occupancy of the condensate state thus neglecting terms corresponding to spontaneous scattering. We checked numerically that these terms do not affect the obtained results.
\begin{figure}[t]
\includegraphics[width=1.\linewidth]{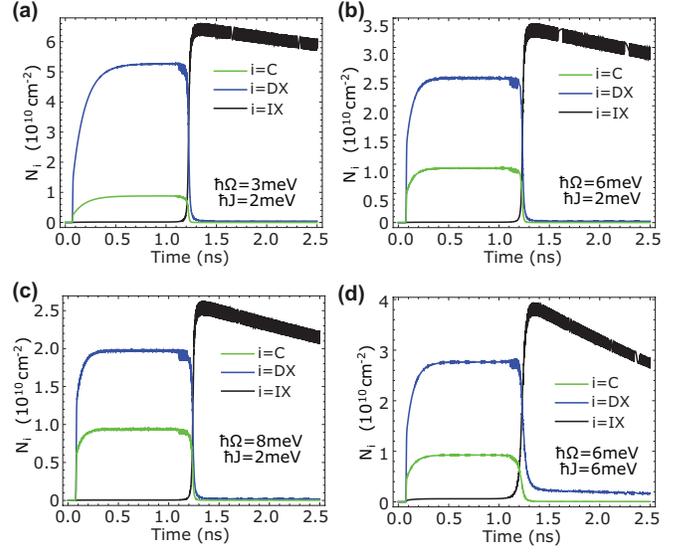}
\caption{(Color online) Transition dynamics shown for different coupling parameters $\Omega$ and $J$. (a, b, c): Tunneling coupling is fixed to $\hbar J = 2$ meV ($L_b = 8$ nm), while Rabi splitting is equal to $\hbar \Omega = 3$ meV (a), $\hbar \Omega = 6$ meV (b), and $\hbar \Omega = 8$ meV (c). (d): Occupation transfer in the system with equal couplings, $\hbar J = \hbar \Omega = 6$ meV.
}
\label{Fig3}
\end{figure}
%


\section{Results and discussion}
We simulate the dipolariton system based on
In$_{0.1}$Ga$_{0.9}$As/GaAs/In$_{0.08}$Ga$_{0.92}$As
heterostructure.\cite{Cristofolini2012} The coupling parameters are
chosen as $\hbar\Omega=3$ meV and $\hbar J=1$ meV. The latter
corresponds to DQW with barrier width of $L_b = 10$ nm. Lifetimes of
the modes are chosen as $\tau_C = 20$ ps, $\tau_{DX}= 1$ ns,
$\tau_{IX} = 100$ ns. The damping rate of reservoir states $\gamma_R
= 1/\tau_{R}$ is defined by lifetime $\tau_R = 0.1$ ns. Initially,
at zero applied field energies of the modes system are tuned to
$\hbar\omega_{DX}- \hbar\omega_{C}=-3$ meV, $\hbar\omega_{IX}^{(0)}
- \hbar\omega_{C}=15$ meV, and we set the reference point $\hbar
\omega_{C} = 0$ without loss of generality [see Fig. \ref{Fig2}(a)
for $t \rightarrow 0$]. The reservoir scattering rate
$W=1/\tau_{sc}$ is defined by characteristic exciton-phonon
scattering time $\tau_{sc}$, and chosen as $200$ ps.\cite{Tignon,
Savona} The temperature of the sample was assumed to be $T = 0.7$ K, being typical for experiments with cold indirect excitons.\cite{Butov2004a}

We start switching on incoherent pump $P(t)$ gradually [Fig.
\ref{Fig2}(a)], populating reservoir exciton states. Due to phonon-exciton interaction particles from
reservoir thermalize and scatter to lowest energy state of lower
dipolariton branch during several hundreds of picoseconds, where the
steady-state is achieved approximately at $0.5$ ns [Fig.
\ref{Fig2}(b)]. By this moment the DX mode is highly populated,
while IX mode remains empty due to large separation in energy and
smallness of interaction with reservoir. The crucial idea now is to
change the bias applied to the system. The IX mode energy then
varies linearly up to large red-detuned value [$\hbar\omega_{IX}^{(\infty)} = -15$ meV], while direct exciton and cavity photon modes energies remain unchanged. The exact shape of IX energy dependence is plotted in Fig. \ref{Fig2}(a), and is described by formula $\hbar\omega_{IX}(t)=\left(\hbar\omega_{IX}^{(0)}-\hbar\omega_{IX}^{(\infty)}\right)/\left(1+\exp^{[t-\tau]/\Delta\tau}\right)+\hbar\omega_{IX}^{(\infty)}$, where parameters are $\tau = 1200$ ps and $\Delta\tau = 50$ ps.

If one changes the bias slowly, the Landau-Zener type transition
between bosonic modes takes place. Namely, the system adiabatically
follows the lower dipolariton branch, and at the final stage of
complete swap gains $99.9\%$ indirect exciton fraction. Alias, in
the basis of bare modes the IX population largely increases due to
transfer from cavity photon and direct exciton (polariton) modes,
where their occupations drop to zero [Fig. \ref{Fig2}(b), after 1.3
ns]. This corresponds to conversion of macroscopically coherent
polariton population to a gas of cold indirect excitons, which
inherit coherence properties and demonstrate long lifetime
(characteristic decay time $>10$ ns for chosen parameters).

We note that successive preparation of cold exciton BEC requires simultaneous switching off the pump $P(t)$, leading to the rapid devastation of the reservoir mode. The following allows to keep high transfer fidelity, where overall particle occupation remains large during the transfer event, but prevents refilling of polaritonic states.
\begin{figure}
\includegraphics[width=1.\linewidth]{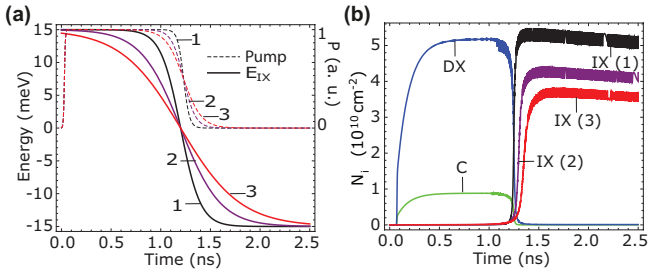}
\caption{(Color online) IX mode dynamics for different adiabaticity parameters. (a): Time dependence of IX mode energy (solid curves, left scale) and pump intensity (dashed curves, right scale). Three different regimes correspond to switching times $\Delta\tau_{1,2,3} = 100, 200, 300$ ps. (b): Occupation number dynamics of the modes shown for different switching regimes. The dynamics of C and DX modes remains unchanged, while the final occupation of IX mode strongly depends on switching parameters.}
\label{Fig4}
\end{figure}

To characterize the system and find optimal parameters for cold
exciton condensate preparation, we proceed considering different
coupling parameters of dipolariton setup. In Fig. \ref{Fig3} the
transition process is demonstrated for several values of Rabi
frequency $\Omega$ and tunneling splitting $J$. First, we fix
tunneling constant to $\hbar J = 2$ meV (as in Fig. \ref{Fig2}) and
vary the strength of light-matter interaction. We find that increase
of Rabi frequency $\Omega$ causes the reduction of transition
efficiency, which can be clearly observed in Figs. \ref{Fig3}(a),
(b), and (c). First, this can be linked to increase of cavity photon
admixture in the lower dipolariton mode, and consequent enlargement
of decay. Second, the change of coupling $\Omega$ in general
strongly influences Landau-Zener transition in three-mode system.

In Fig. \ref{Fig3}(d) we show the population transfer for the case of equal couplings $\hbar J= \hbar \Omega = 6$ meV, corresponding to the sample discussed in the Ref. [46]. We observe that population transfer for these parameters is not perfect, and thus conclude that $J < \Omega$ condition shall be followed. Furthermore, we note that large values of tunneling constant $J$ lead to stronger mixing of IX and DX modes, with consequent decrease of the lower dipolariton lifetime after switching.

Next, we study the influence of electric field switching process on the performance of population transfer. The main parameter here is a switching time $\tau$, which also determines the switching rate $c = \Delta \tau^{-1}/4$ at which energy linearly decreases as $\sim-c t$ around transition point $\tau$. It defines the adiabaticity of transition, and is of high importance for successive Landau-Zener transition. In particular, the optimal transfer requires slow variation of detuning, as compared to the energy distance between modes in the anticrossing point of dressed modes.

In Fig. \ref{Fig4} the time dynamics of the system is presented for
different switching rates. We considered three regimes, altering
both the rate of detuning switching and the front of pump switching,
in order to keep density of the particles at the same level [Fig.
\ref{Fig4}(a)]. We find that the voltage switching rate almost does
not affect on the polariton population at steady-state condition,
but strongly influences the efficiency of the transition, defining
the IX mode occupation [Fig. \ref{Fig4}(b)]. For the switching time
$\Delta \tau_1 = 100$ ps nearly perfect transfer was achieved. This
confirms that the characteristic time corresponding to anticrossing
energy distance is $\sim 1$ ps, and transition is adiabatic. However, decreasing the switching rate to $\Delta \tau_2 = 200$ ps and $\Delta \tau_3 =300$ ps we revealed the reduction of transfer efficiency, signifying of an existence of an optimal rate $\Delta \tau$. The process of spoiling of transition is related to open-dissipative nature of the system under the study.

\begin{figure}[t]
\includegraphics[width=1.\linewidth]{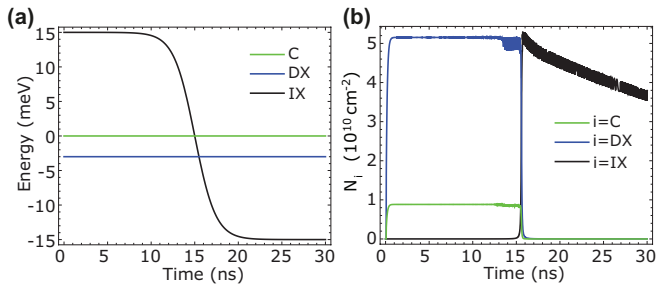}
\caption{(Color online) (a): Time dependence of energy of the modes for modified dipolariton system, where ultra-slow change of detuning is implemented. (b): Population transfer in dipolariton system for ultra-slow detuning change. Parameters of the calculation are: $\tau=15$ ns, $\Delta\tau=1.2$ ns.}
\label{Fig5}
\end{figure}

Finally, we acknowledge the presence of another bound which puts
limitation on detuning switching rate. It comes from the
experimental limitation of DC voltage sweep rate, which typically
cannot outperform gigahertz repetition rates. While state-of-the-art
devices with fast switching are currently engineered,\cite{Tonouchi2007} we probe the
possibility to decrease the sweep rate up to $100$ MHz range. In
Fig. \ref{Fig5} we show that the transition can take place for
switching time being in the tens of nanoseconds range [Fig.
\ref{Fig5}(a)], easily achievable with current technologies. The
transition efficiency in this case drops as opposed to
less-than-nanosecond switching time. At the same time, a decent
population of indirect exciton gas can be achieved for certain
pumping conditions, which persists for tens of nanoseconds. The
parameters of the system for this calculation were modified to
$\hbar J = 0.6$ meV and $\hbar \Omega = 3$ meV.

Finally, let us discuss the immediate consequences of proposed
scheme for cold exciton gas preparation. Being based on conversion
of optically created polaritons to an indirect exciton gas, it
ensures the preservation of the spatial shape of initial cloud, and
does not involve separate injection of electron and hole carriers.
This is in contrast to typical excitation scheme in the indirect
exciton experiments, where optical generation of holes and
electrical injection of electrons is
used.\cite{Butov2004a,Butov2007} The following allows to test the
possible explanation of IX ring appearance based on the
electrostatic reasoning.\cite{Butov2004b,Rapaport2004}

Moreover, we note that our proposal can be tested in first
approximation even without the presence of an optical
microresonator. In this case only coupled DX and IX modes are
considered, and conversion of optically active direct excitons to a
cloud of indirect excitons can be realized. However, on the contrary
to the full dipolaritonic setup, no condensation effects for direct
excitons are expected which can limit the efficiency of the
proposed protocol.


\section{Conclusions}

In conclusion, we proposed the way for an on-demand optical
preparation of a cold exciton condensate based on Landau-Zener
bosonic transfer in a dipolariton system. The protocol is based on
several steps. First stage corresponds to initial preparation of
polariton condensate with high cavity photon and direct exciton
fractions, while indirect exciton mode is located high in energy at
zero external voltage. Next, applying electric field the IX energy
is lowered to far red-detuned value, where adiabatic following of
the lower dipolariton mode converts particles to indirect excitons
with inherited coherence properties. Finally, to reduce residual
effects of cavity an optical incoherent pump of polaritonic
reservoir states shall be switched off during the transfer event. We
analyzed the population transfer for various sets of parameters and
switching conditions, and demonstrated that adiabatic cold exciton
preparation is experimentally feasible in currently existing setups.

\section*{Acknowledgements}
We thank S. A. Tarasenko, S. Morina, and H. Sigurdsson for the
useful discussions. The work was supported by the Tier 1 project
“Polaritons for novel device applications,” FP7 IRSES projects
POLAPHEN, POLATER and QOCaN, and FP7 ITN NOTEDEV network.



\begin{thebibliography}{99}

\bibitem{Pitaevskii2006} L. Pitaevskii and S. Stringari, \textit{Bose-Einstein Condensation} (Oxford University Press, Oxford, 2003).

\bibitem{Anderson1995} M. H. Anderson, J. R. Ensher, M. R. Matthews, C. E. Wieman, and E. A. Cornell, Science \textbf{269}, 198 (1995).

\bibitem{Davis1995} K. B. Davis, M.-O. Mewes, M. R. Andrews, N. J. van Druten, D. S. Durfee, D. M. Kurn, and W. Ketterle, Phys. Rev. Lett. \textbf{75}, 3969 (1995).

\bibitem{Keldysh1965} L. V. Keldysh and Yu. V. Kopaev, \emph{Sov. Phys. Solid State} \textbf{6}, 2219 (1965); L. V. Keldysh and A. N. Kozlov, Sov. Phys. JETP \textbf{27}, 521 (1968).

\bibitem{Blatt1962} J. M. Blatt, K. W. B\"{o}er, and W. Brandt, Phys. Rev. \textbf{126}, 1691 (1962).

\bibitem{SnokeBook} S. A. Moskalenko and D. W. Snoke, \textit{Bose-Einstein Condensation of Excitons and Biexcitons} (Cambridge University Press, Cambridge, U.K., 2000).

\bibitem{Kavokin2007} A. V. Kavokin, J. J. Baumberg, G. Malpuech, and F. P. Laussy, {\it Microcavities} (Oxford University Press, New York, 2007).

\bibitem{Deng2010} H. Deng, H. Haug, and Y. Yamamoto, Rev. Mod. Phys. 82, 1489 (2010).

\bibitem{Kasprzak2006} J. Kasprzak, M. Richard, S. Kundermann, A. Baas, P. Jeambrun, J. M. J. Keeling, F. M. Marchetti, M. H. Szymanska, R. Andre, J. L. Staehli, V. Savona, P. B. Littlewood, B. Deveaud, and Le Si Dang, Nature (London) \textbf{443}, 409 (2006).

\bibitem{Balili2007} R. Balili, V. Hartwell, D. Snoke, L. Pfeiffer, and K. West, Science \textbf{316}, 1007 (2007).

\bibitem{Wertz2010} E. Wertz, L. Ferrier, D. D. Solnyshkov, R. Johne, D. Sanvitto, A. Lema\^{i}tre, I. Sagnes, R. Grousson, A. V. Kavokin, P. Senellart, G. Malpuech, and J. Bloch, Nature Phys. \textbf{6}, 860 (2010).

\bibitem{Fischer2014} J. Fischer, S. Brodbeck, A. V. Chernenko, I. Lederer, A. Rahimi-Iman, M. Amthor, V. D. Kulakovskii, L. Worschech, M. Kamp, M. Durnev, C. Schneider, A. V. Kavokin, and S. H\"{o}fling, Phys. Rev. Lett. \textbf{112}, 093902 (2014).

\bibitem{Baumberg2008} J. J. Baumberg, A. V. Kavokin, S. Christopoulos, A. J. D. Grundy, R. Butt\'{e}, G. Christmann, D. D. Solnyshkov, G. Malpuech, G. Baldassarri H\"{o}ger von Hogersthal, E. Feltin, J.-F. Carlin, and N. Grandjean, Phys. Rev.  Lett. \textbf{101}, 136409 (2008).

\bibitem{Plumhof2014} J. D. Plumhof, T. St\"{o}ferle, L. Mai, U. Scherf, and R. F. Mahrt, Nature Mater. \textbf{13}, 247 (2014) .
%

\bibitem{Schneider2013} C. Schneider, A. Rahimi-Iman, Na Young Kim, J. Fischer, I. G. Savenko, M. Amthor, M. Lermer, A. Wolf, L. Worschech, V. D. Kulakovskii, I. A. Shelykh, M. Kamp, S. Reitzenstein, A. Forchel, Y. Yamamoto, and S. Hofling, Nature (London) \textbf{497}, 348 (2013).

\bibitem{Bhattacharya2013} P. Bhattacharya, B. Xiao, A. Das, S. Bhowmick, and J. Heo, Phys. Rev. Lett. \textbf{110}, 206403 (2013).

\bibitem{Sich2013} M. Sich, D. N. Krizhanovskii, M. S. Skolnick, A. V. Gorbach, R. Hartley, D. V. Skryabin, E. A. Cerda-M\'{e}ndez, K. Biermann, R. Hey, and P. V. Santos, Nature Photon. \textbf{6}, 50 (2012).

\bibitem{Sich2014} M. Sich, F. Fras, J. K. Chana, M. S. Skolnick, D. N. Krizhanovskii, A. V. Gorbach, R. Hartley, D. V. Skryabin, S. S. Gavrilov, E. A. Cerda-M\'{e}ndez, K. Biermann, R. Hey, and P. V. Santos, Phys. Rev. Lett. \textbf{112}, 046403 (2014).

\bibitem{Amo2011} A. Amo, S. Pigeon, D. Sanvitto, V. G. Sala, R. Hivet, I. Carusotto, F. Pisanello, G. Lem\'{e}nager, R. Houdr\'{e}, E Giacobino, C. Ciuti, and A. Bramati, Science \textbf{332}, 1167 (2011).

\bibitem{Lagoudakis2008} K. G. Lagoudakis, M. Wouters, M. Richard, A. Baas, I. Carusotto, R. Andr\'{e}, Le Si Dang, and B. Deveaud-Pl\'{e}dran, Nature Phys. \textbf{4}, 706 (2008).

\bibitem{Manni2012} F. Manni, K. G. Lagoudakis, T. C. H Liew, R. Andr\'{e}, V. Savona, and B. Deveaud, Nature Commun. \textbf{3}, 1309 (2012); doi:10.1038/ncomms2310.

\bibitem{Leyder2007} C. Leyder, M. Romanelli, J. Ph. Karr, E. Giacobino, T. C. H. Liew, M. M. Glazov, A. V. Kavokin, G. Malpuech, and A. Bramati, Nature Phys. \textbf{3}, 628 (2007).

\bibitem{Tosi2011} G. Tosi, F. M. Marchetti, D. Sanvitto, C. Ant\'{o}n, M. H. Szyma\'{n}ska, A. Berceanu, C. Tejedor, L. Marrucci, A. Lema\^{i}tre, J. Bloch, and L. Vina, Phys. Rev. Lett. \textbf{107}, 036401 (2011).

\bibitem{Adrados2010} C. Adrados, A. Amo, T. C. H. Liew, R. Hivet, R. Houdr\'{e}, E. Giacobino, A. V. Kavokin, and A. Bramati, Phys. Rev. Lett. \textbf{105}, 216403 (2010).

\bibitem{Shelykh2010} I. A. Shelykh, A. V. Kavokin, Y. G. Rubo, T. C. H. Liew, and G. Malpuech, Semicond. Sci. Technol. \textbf{25}, 013001 (2010).

\bibitem{Espinosa2013} T. Espinosa-Ortega and T. C. H. Liew, Phys. Rev. B \textbf{87}, 195305 (2013).

\bibitem{Liew2010} T. C. H. Liew, A. V. Kavokin, T. Ostatnick\'{y}, M. Kaliteevski, I. A. Shelykh, and R. A. Abram, Phys. Rev. B \textbf{82}, 033302 (2010).

\bibitem{CarusottoRev} I. Carusotto and C. Ciuti, Rev. Mod. Phys. \textbf{85}, 299 (2013).






\bibitem{High2012} A. A. High, J. R. Leonard, A. T. Hammack, M. M. Fogler, L. V. Butov, A. V. Kavokin, K. L. Campman, and A. C. Gossard, Nature (London) \textbf{483}, 584 (2012).

\bibitem{Butov2004a} L. V. Butov, J. Phys.: Condens. Matter \textbf{16}, R1577 (2004).

\bibitem{Butov2007} L. V. Butov, J. Phys. Condens. Matter \textbf{19}, 295202 (2007).

\bibitem{Butov2002} L. V. Butov, A. C. Gossard, and D. S. Chemla, Nature \textbf{418}, 751 (2002).

\bibitem{Snoke2002} D. Snoke, S. Denev, Y. Liu, L. Pfeiffer, and K. West, Nature \textbf{418}, 754 (2002).

\bibitem{Dubin2012} M. Alloing, D. Fuster, Y. Gonz\'{a}lez, L. Gonz\'{a}lez, and F. Dubin, arXiv:1210.3176 (2012).

\bibitem{High2013} A. A. High, A. T. Hammack, J. R. Leonard, Sen Yang, L. V. Butov, T. Ostatnick\'{y}, M. Vladimirova, A. V. Kavokin, T. C. H. Liew, K. L. Campman, and A. C. Gossard, Phys. Rev. Lett. \textbf{110}, 246403 (2013).

\bibitem{Leonard2009} J. R. Leonard, Y. Y. Kuznetsova, Sen Yang, L. V. Butov, T. Ostatnick\'{y}, A. Kavokin, and A. C. Gossard, Nano Lett. \textbf{9}, 4204 (2009).

\bibitem{Violante2014} A. Violante, R. Hey, and P. V. Santos, arXiv:1408.4547 (2014).

\bibitem{Winbow2011} A. G. Winbow, J. R. Leonard, M. Remeika, Y. Y. Kuznetsova, A. A. High, A. T. Hammack, L. V. Butov, J. Wilkes, A. A. Guenther, A. L. Ivanov, M. Hanson, and A. C. Gossard, Phys. Rev. Lett. \textbf{106}, 196806 (2011).

\bibitem{Andreakou2014} P. Andreakou, S. V. Poltavtsev, J. R. Leonard, E. V. Calman, M. Remeika, Y. Y. Kuznetsova, L. V. Butov, J. Wilkes, M. Hanson, and A. C. Gossard, Appl. Phys. Lett. \textbf{104}, 091101 (2014).

\bibitem{Lazic2014} S. Lazi\'{c}, A. Violante, K. Cohen, R. Hey, R. Rapaport, and P. V. Santos, Phys. Rev. B \textbf{89}, 085313 (2014).

\bibitem{Sivalertporn2012} K. Sivalertporn, L. Mouchliadis, A. L. Ivanov, R. Philp, and E. A. Muljarov, Phys. Rev. B \textbf{85}, 045207 (2012).

\bibitem{Butov2001} L. V. Butov, A. L. Ivanov, A. Imamoglu, P. B. Littlewood, A. A. Shashkin, V. T. Dolgopolov, K. L. Campman, and A. C. Gossard, Phys. Rev. Lett. \textbf{86}, 5608 (2001).

\bibitem{Nalitov2014a} A. V. Nalitov, M. Vladimirova, A. V. Kavokin, L. V. Butov, and N. A. Gippius, Phys. Rev. B \textbf{89}, 155309 (2014).

\bibitem{Butov2004b} L. V. Butov, L. S. Levitov, A. V. Mintsev, B. D. Simons, A. C. Gossard, and D. S. Chemla, Phys. Rev. Lett. \textbf{92}, 117404 (2004).
%

\bibitem{Rapaport2004} R. Rapaport, Gang Chen, D. Snoke, Steven H. Simon, Loren Pfeiffer, Ken West, Y. Liu, and S. Denev, Phys. Rev. Lett. \textbf{92}, 117405 (2004).
%


\bibitem{Cristofolini2012} P. Cristofolini, G. Christmann, S. I. Tsintzos, G. Deligeorgis, G. Konstantinidis, Z. Hatzopoulos, P. G. Savvidis, and J. J. Baumberg, Science \textbf{336}, 704 (2012).

\bibitem{Christmann2011} G. Christmann, A. Askitopoulos, G. Deligeorgis, Z. Hatzopoulos, S. I. Tsintzos, P. G. Savvidis, and J. J. Baumberg, Appl. Phys. Lett. \textbf{98}, 081111 (2011).

\bibitem{Kyriienko2013} O. Kyriienko, A. V. Kavokin, and I. A. Shelykh, Phys. Rev. Lett. \textbf{111}, 176401 (2013).

\bibitem{Kristinsson2013} K. Kristinsson, O. Kyriienko, T. C. H. Liew, and I. A. Shelykh, Phys. Rev. B \textbf{88}, 245303 (2013).

\bibitem{Kristinsson2014} K. Kristinsson, O. Kyriienko, and I. A. Shelykh, Phys. Rev. A \textbf{89}, 023836 (2014).

\bibitem{Kyriienko2014} O. Kyriienko, I. A. Shelykh, and T. C. H. Liew, Phys. Rev. A \textbf{90}, 033807 (2014).

\bibitem{Nalitov2014b} A. Nalitov, D. Solnyshkov, N. Gippius, and G. Malpuech, Proceeding of PLMCN2014, Montpellier, France.

\bibitem{Sinitsyn2013} N. A. Sinitsyn, Phys. Rev. A \textbf{87}, 032701 (2013).

\bibitem{Liew2012} T. C. H. Liew and I. A. Shelykh, J. Phys. B: At. Mol. Opt. Phys. \textbf{45}, 245003 (2012).

\bibitem{Wouters2007} M. Wouters and I. Carusotto, Phys. Rev. Lett. \textbf{99}, 140402 (2007).

\bibitem{SM} See Supplemental Material [online] for the details of derivation.

\bibitem{Tignon} S. Huppert, O. Lafont, E. Baudin, J. Tignon and R. Ferreira, arXiv:1409.2780v1.

\bibitem{Savona} F. Tassone, C. Piermarocchi, V. Savona, A. Quattropani, and P. Schwendimann, Phys.
Rev. B \textbf{53}, R7642 (1996).

\bibitem{Tonouchi2007} M. Tonouchi, Nature Photonics \textbf{1}, 97 (2007).

\end{thebibliography}
\end{document}


\title{Supplemental Material: Adiabatic preparation of a cold exciton condensate}

\author{V. Shahnazaryan}
\affiliation{Science Institute, University of Iceland, Dunhagi-3, IS-107, Reykjavik, Iceland}

\author{O. Kyriienko}
\affiliation{Science Institute, University of Iceland, Dunhagi-3, IS-107, Reykjavik, Iceland}
\affiliation{Division of Physics and Applied Physics, Nanyang Technological University 637371, Singapore}
\affiliation{QUANTOP, Danish Quantum Optics Center, Niels Bohr Institute, University of Copenhagen, Blegdamsvej 17, DK-2100 Copenhagen, Denmark}

\author{I. A. Shelykh}
\affiliation{Science Institute, University of Iceland, Dunhagi-3, IS-107, Reykjavik, Iceland}
\affiliation{Division of Physics and Applied Physics, Nanyang Technological University 637371, Singapore}

\date{\today}

\maketitle

\appendix
\section{Equations of motion in first order mean field theory}

Here we present the derivation of equations for time dynamics of the system [Eqs. (4)-(7) in the main text].

\textit{Coherent part}. The coherent part of dynamics for cavity photon occupation number $\langle\hat{a}\rangle$ can be found as
\begin{align}
&\left. \frac{\partial\langle\hat{a}\rangle}{\partial t}\right|_{coh}=\frac{i}{\hbar}Tr\{\hat{a} [\rho,\hat{H}_{coh}] \}=\frac{i}{\hbar}Tr\{\rho[\hat{H}_{coh},\hat{a}]\}\\ \notag
&=\frac{i}{\hbar}Tr\{\rho(-\hbar\omega_C\hat{a}-\frac{\hbar\Omega}{2}\hat{b})\}=-i\omega_C\langle\hat{a}\rangle-i\frac{\Omega}{2}\langle\hat{b}\rangle.
\end{align}
In the same fashion for expectation values of operators $\hat{b}$, $\hat{c}$ we have
\begin{equation}
\left. \frac{\partial\langle\hat{b}\rangle}{\partial t}\right|_{coh}=-i\omega_{DX}\langle\hat{b}\rangle-i\frac{\Omega}{2}\langle\hat{a}\rangle+i\frac{J}{2}\langle\hat{c}\rangle,
\end{equation}
\begin{equation}
\left. \frac{\partial\langle\hat{c}\rangle}{\partial t}\right|_{coh}=-i\omega_{IX}\langle\hat{c}\rangle+i\frac{J}{2}\langle\hat{b}\rangle.
\end{equation}

\textit{Dissipation}. To describe the incoherent processes the master equation approach with dissipators written in Lindblad form can be used.\cite{Kyriienko2013} Particularly, the decay of cavity mod can be treated as
\begin{align}
\left. \frac{\partial\langle\hat{a}\rangle}{\partial t}\right|_{dis}&=Tr\{\hat{a}\hat{\mathcal{L}}_{\hat{a}}^{(dis)}\rho\}=Tr\{\hat{a}\frac{\gamma_C}{2}(2\hat{a}\rho\hat{a}^\dag-\hat{a}^\dag\hat{a}\rho-\rho\hat{a}^\dag\hat{a})\} \notag \\
&=\frac{\gamma_C}{2}Tr\{2\rho\hat{a}^\dag\hat{a}\hat{a}-\rho\hat{a}\hat{a}^\dag\hat{a}-\rho\hat{a}^\dag\hat{a}\hat{a}\}=-\frac{\gamma_C}{2}\langle\hat{a}\rangle.
\end{align}
Similarly,
\begin{align}
&\left. \frac{\partial\langle\hat{b}\rangle}{\partial t}\right|_{dis}=-\frac{\gamma_{DX}}{2}\langle\hat{b}\rangle, \notag \\
&\left. \frac{\partial\langle\hat{c}\rangle}{\partial t}\right|_{dis}=-\frac{\gamma_{IX}}{2}\langle\hat{c}\rangle,\\
&\left. \frac{\partial n_R}{\partial t}\right|_{dis}=-\gamma_{R} n_R. \notag
\end{align}

\textit{Interaction with reservoir}. The dynamics of the system caused by interaction with exciton reservoir can be calculated in Born-Markov approximation as
\begin{equation}
\left. \frac{\partial\langle\hat{b}\rangle}{\partial t}\right|_{int}=\delta_{\Delta E} \left\{\left\langle\left[\hat{H}^-,[\hat{b},\hat{H}^+]\right]\right\rangle+\left\langle\left[\hat{H}^+,[\hat{b},\hat{H}^-]\right]\right\rangle\right\},
\end{equation}
where $\delta_{R}$ is inverse broadening of exciton states divided by $\hbar^2$. Calculation of commutators one by one gives $[\hat{b},\hat{H}^-]=0$, $[\hat{b},\hat{H}^+]=R_{ph}\sum_k\hat{r}_k\hat{d}^+_k$, $\left[\hat{H}^+,[\hat{b},\hat{H}^-]\right]=R_{ph}^2\sum_k(\hat{r}^\dag_k\hat{r}_k-\hat{d}_k^\dag\hat{d}_k)\hat{b}$. Finally,
\begin{align}
&\left. \frac{\partial\langle\hat{b}\rangle}{\partial t}\right|_{int}=\delta_{\Delta E}R_{ph}^2\sum_k(n_k^R-n_k^{ph})\langle\hat{b}\rangle \\ \notag
&=\delta_{\Delta E}R_{ph}^2 \left(N_R-\sum_k n_k^{ph}\right)\langle\hat{b}\rangle.
\end{align}

In the same way for reservoir states we get
\begin{align}
&\left.\frac{\partial N_R}{\partial t}\right|_{int}=\sum_k\left. \frac{\partial n_k^R}{\partial t}\right|_{int}\\
&=\sum_k\left\{\left\langle\left[\hat{H}^-,[\hat{r}^\dag\hat{r},\hat{H}^+]\right]\right\rangle+\left\langle\left[\hat{H}^+,[\hat{r}^\dag\hat{r},\hat{H}^-]\right]\right\rangle\right\} \notag \\
=&-2\delta_{\Delta E}R_{ph}^2\left(|\langle\hat{b}\rangle|^2 N_R+\sum_k[(n_k^{ph}+1)n^R_k-|\langle\hat{b}\rangle|^2n_k^{ph}]\right). \notag
\end{align}
Here the relevant term corresponds to the stimulated scattering between reservoir and coherent mode, while spontaneous scattering can be usually neglected.

\textit{Incoherent pump}. The pump of reservoir can be introduced using the Lindblad superoperator for incoherent pumping of each reservoir state with momentum $q$, and is given by:\cite{Ostrovskaya2014}
\begin{align}
&\left.\frac{\partial n_q^R}{\partial t}\right|_{pump}=Tr\left\{\hat{r}^\dag_q\hat{r}_q \sum_k P_k(t)\left(\hat{r}_k\rho\hat{r}_k^\dagger +\hat{r}^\dagger_k\rho\hat{r}_k- \hat{r}_k^\dagger\hat{r}_k\rho \right.\right. \\ \notag &\left.\left.-\rho\hat{r}_k\hat{r}_k^\dagger \right) \right\} = P_{q}(t),
\end{align}
where $P_{q}(t)$ corresponds to an incoherent pumping rate of a single reservoir state. Considering identical reservoir states, we can write equation for the total reservoir occupation as
\begin{align}
&\left.\frac{\partial N_R}{\partial t}\right|_{pump}=\sum_q\left. \frac{\partial n_q^R}{\partial t}\right|_{pump} = \sum_q P_q(t) \equiv P(t),
\end{align}
where we defined the total incoherent pumping rate $P(t)$ as a sum of single state pumps.

%
%

Combining the all contributions, for the time dynamics we come to the system of coupled equations (4)-(7).
